\documentclass[12pt]{article}
\usepackage{amsmath}
\usepackage{graphics,graphicx,epsfig}
\usepackage{amssymb}
\usepackage{appendix}

\usepackage{caption}
\usepackage{subcaption}

\usepackage{color}

\newcommand{\be}{\begin{equation}}
\newcommand{\ee}{\end{equation}}
\newcommand{\bea}{\begin{eqnarray}}
\newcommand{\eea}{\end{eqnarray}}
\newcommand{\beal}{\begin{aligned}}
\newcommand{\eeal}{\end{aligned}}

\def\varep{\varepsilon}
\def\N{ {\cal N}_\omega }
\def\hb{ {\cal H}_b }
\def\hc{ {\cal H}_c }

\def\mbar{m_N}

\begin{document}

\begin{titlepage}
\vfill
\begin{flushright}
ACFI-T19-07
\end{flushright}

\vfill
\begin{center}
\baselineskip=16pt
{\Large\bf Schottky Anomaly of deSitter Black Holes}

\vskip 10.mm

{\bf   Jack Dinsmore$^{(a)}$, Patrick Draper$^{(b)}$, David Kastor$^{(a)}$, Yue Qiu$^{(a)}$, \\
and Jennie Traschen$^{(a)}$} 

\vskip 0.4cm

{$^{(a)}$ Amherst Center for Fundamental Interactions}\\
{Department of Physics, University of Massachusetts, Amherst, MA 01003, USA}\\
{$^{(b)}$ Department of Physics, University of Illinois, Urbana, IL 61801}\\
\vspace{0.3cm}
\vskip 4pt

\vskip 0.1 in Email: \texttt{jtdinsmo@mit.edu,  pdraper@illinois.edu, kastor@physics.umass.edu, yqiu@physics.umass.edu, traschen@physics.umass.edu}
\vspace{6pt}
\end{center}
\vskip 0.3in
\par
\begin{center}
{\bf Abstract}  
\end{center}
\begin{quote}
The interplay of black hole and cosmological horizons introduces distinctive thermodynamic behavior for deSitter black holes, including well-known upper bounds for the mass and entropy.  We point to a new such feature, a Schottky peak in the heat capacity of Schwarzschild-deSitter (SdS) black holes.  With this behavior in mind, we explore statistical models for the underlying quantum degrees of freedom of SdS holes.  While a simple two-state spin model gives Schottky behavior, in order to capture the non-equilibrium nature of the SdS system we consider a system with a large number of non-interacting spins.  We examine to what extent constrained states of this system  reproduce the thermodynamic properties of the black hole.  We also review results of a recent study of particle production in SdS spacetimes in light of the Schottky anomaly and our spin models.
\vfill
\vskip 2.mm
\end{quote}
\end{titlepage}

\section{Introduction}

Black hole thermodynamics provides a window into the quantum character of spacetime, which remains experimentally and observationally inaccessible.  The field has been deepened  through the gauge-gravity correspondence \cite{Maldacena:1997re}, where thermal properties of black holes map onto phenomena in dual quantum field theories.  A prominent example in this context is the Hawking-Page transition of AdS black holes  \cite{Hawking:1982dh}, which maps to the deconfining phase transition in Supersymmetric Yang-Mills theory \cite{Witten:1998qj}. The deSitter counterpart \cite{Strominger:2001pn} of AdS/CFT is considerably less understood, but here too distinctive thermal features of deSitter black holes may map to interesting behavior in a dual theory.  

Recent interest has also focused on an ``extended phase space''  for (A)dS black holes, in which the cosmological constant, interpreted as a pressure, is considered to be a thermodynamic variable with a conjugate thermodynamic volume. 
An extended first law, including variations in $\Lambda$, was proved in \cite{Kastor:2009wy}, and the study of black hole phase transitions in the extended phase space for AdS black holes was initiated in \cite{Kubiznak:2012wp} (see \cite{Kubiznak:2016qmn} for a review and many additional references).  The extended phase space for dS black holes was studied in \cite{Dolan:2013ft}.


Our focus in this paper will be on the thermodynamics of dS black holes and, in particular, on what they may tell us about underlying quantum degrees of freedom.
Black hole spacetimes with $\Lambda>0$ have both  black hole and  cosmological horizons, which generically
radiate at different temperatures. The entropy of each horizon is given by one-quarter its area, and
the total entropy of the spacetime is equal to the sum\footnote{This total entropy for deSitter black holes has been shown to be physically relevant in a number of different settings.  For example, the analogue of BPS black holes in deSitter are RNdS spacetimes with $|Q|=M$.  For these spacetimes, the black hole and cosmological horizon temperatures are equal and the total spacetime entropy is maximized \cite{Kastor:1992nn,Kastor:1993mj}.  The sum of the (Euclidean) horizon areas for SdS black holes also determines the rate of false vacuum decay in the presence of black hole impurities \cite{Gregory:2013hja}.}
\begin{equation}\label{totentropy}
S= {1\over 4} \left( A_b + A_c  \right),
\end{equation}
Like the $\Lambda =0$ black holes, the black hole temperature decreases monotonically as the mass increases.
However, the presence of the  cosmological horizon introduces marked differences for black holes with positive $\Lambda$ 
from the $\Lambda \leq 0$ cases. In particular, for Schwarzschild-deSitter (SdS) black holes \cite{kottler} there is a maximum mass and a maximum entropy for the black hole. The 
maximum mass occurs in the limit that the two horizons approach each other and both horizon temperatures approach zero.
This turns out to be a configuration of  minimum total entropy.  As the black hole mass increases from zero to its maximum value, with the black hole temperature sweeping from infinity down to zero, the total entropy decreases monotonically from its maximum value in pure deSitter spacetime with no black hole to this minimum.

As we will see below, the evolution of the total entropy as a function of the black hole temperature displays an interesting feature, known in statistical physics as a Schottky anomaly.  Although the total entropy is a monotonically-increasing function of temperature, its rate of increase is not constant.  The total entropy is relatively flat at both high and low temperatures, while increasing steeply for mid-range temperatures, leading to a peak in the plot of its rate of change.  A similar (though inverted) peak occurs in the heat capacity, defined as the rate of change of mass with black hole temperature.  

As emphasized recently \cite{Johnson:2019vqf} the presence of a thermodynamic Schottky peak can provide an important window into properties of underlying the statistical degrees of freedom\footnote{A Schottky anomaly has been previously noted in  black hole physics in reference \cite{Grumiller:2014oha}.}.  For SdS black holes, the existence of the Schottky peak depends on the presence of the cosmological horizon, {\it i.e.} there is no similar peak in the heat capacity for Schwarzschild black holes.  The simplest statistical model displaying Schottky behavior is a two-state spin system.  We will note some conceptual resonance between this system and the horizons of an SdS spacetime.  

However, such a limited system is not adequate to model the out-of-equilibrium nature of the SdS system, with independent temperatures for the two horizons. Some time ago it was suggested by Banks, Fiol, and Morisse that the SdS entropy should be thought of as the entropy of a set of constrained states in a system with a finite Hilbert space~\cite{Banks:2006rx}.\footnote{More generally, Banks and Fischler have argued that any localized excitations in asymptotically flat or de Sitter spacetimes should be understood as constrained, low-entropy states of systems on the horizon~\cite{Banks:2013fr}. These ideas underly the theory of Holographic Space Time (see, for example,~\cite{Banks:2010tj}).}
Following \cite{Banks:2006rx} (see also \cite{Banks:2018jqo,Albrecht:2004ke}), we introduce a picture of SdS black holes as non-equilibrium constrained states of a statistical system composed of a thermal deSitter bath with a black hole subsystem.  We then explore to what extent the thermal properties of SdS spacetimes can be captured within a model with a large number of non-interacting spins.
We further explore the non-equilibrium nature of SdS black holes through a discussion of recent results \cite{qiu} on particle production from the black hole and cosmological horizons in the context of the Schottky anomaly and our spin models for the underlying degrees of freedom.


The plan for the paper is as follows.  We begin in Section (\ref{sdssection}) by presenting the basic properties of Schwarzschild-deSitter (SdS) spacetimes.  
In Section (\ref{bathsection}), we sketch a picture for the underlying quantum states of SdS black holes in terms of non-equilibrium, constrained states of a thermodynamic system.  In Section (\ref{twolevelsec}), we note the presence of a Schottky anomalous peak in the SdS thermal system.  
We also present the simple example of a  two-state statistical model that displays a Schottky peak in its specific heat.  
In Section (\ref{spinsection}), we introduce constrained, non-equilibrium states of a system of a large number of non-interacting spins and explore which features of SdS black holes can be faithfully modeled by this system.  In Section (\ref{quantumsec}), we report relevant results from a study of particle production in SdS spacetimes \cite{qiu} that display features of the out of equilibrium nature of SdS black holes at the quantum level, in order to compare with our models.  Finally, we offer some conclusions and discussion in Section (\ref{conclusions}).

%

\section{Schwarzschild-deSitter spacetimes}\label{sdssection}

\begin{figure}[h]
\begin{center}\includegraphics[width=0.60\textwidth]{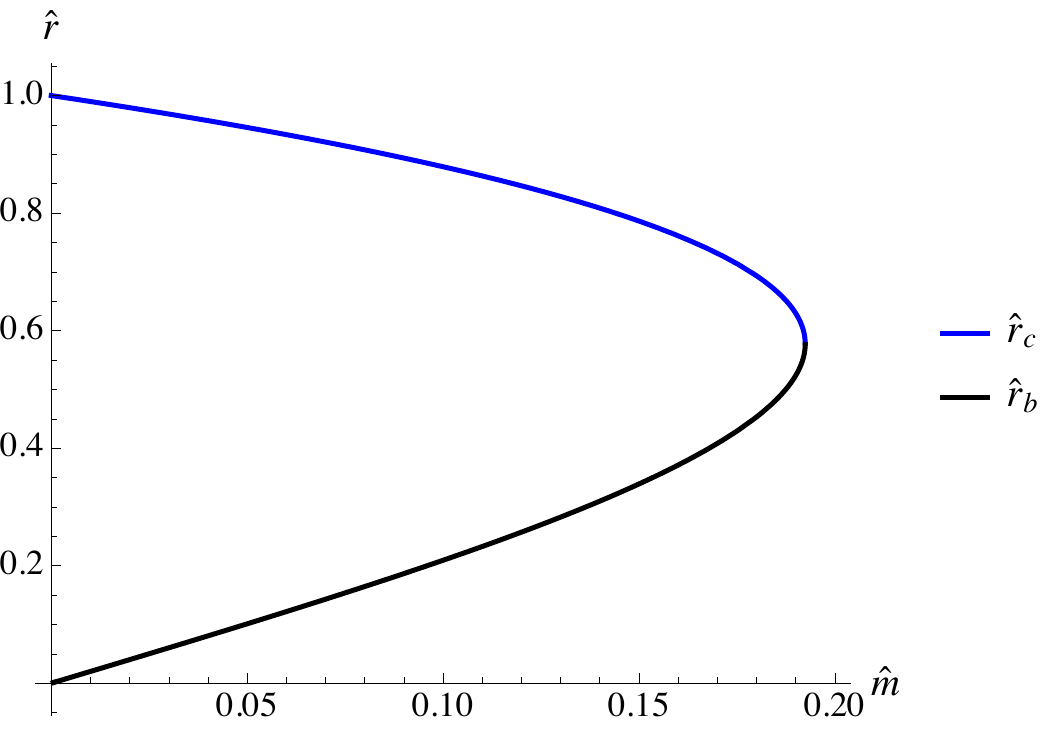}\end{center}
\caption{Plots of normalized black hole and cosmological horizon radii $\hat r_b\equiv r_b/l$ and $\hat r_c=r_c/l$ versus $\hat m=m/l$. For $\hat m=0$, the cosmological horizon radius starts out at its deSitter value $\hat r_c=1$.  As $\hat m$ is increased towards its maximum value $\hat m = 1/(3\sqrt{3})\simeq 0.192$, the two horizon radii come together at the common value $\hat r_b=\hat r_c=1/\sqrt{3}\simeq 0.577$.}  
\label{horizonplot}
\end{figure}

The Schwarzschild-deSitter (SdS) spacetime \cite{kottler} is a solution to Einstein gravity with a positive cosmological constant $\Lambda$. The SdS metric  is parameterized by the mass $m$ and cosmological constant $\Lambda$, or equivalently the deSitter length scale $l$ with $\Lambda = 3/l^2 $,
\begin{equation}\label{SdS}
ds^2 = -f(r)dt^2 +{dr^2\over f(r)} +r^2d\Omega^2,\qquad f(r) =1 - {2m\over r}-{r^2\over l^2}
\end{equation}
%
In the limit $l\rightarrow\infty$, the SdS metric reduces to the Schwarzschild spacetime with black hole horizon at $r_b=2m$, while for $m=0$ it reduces to deSitter spacetime which has a cosmological horizon at $r_c=l$.  The SdS spacetime has regular Killing horizons provided that the mass is in the range
%
\begin{equation}\label{range}
0\le m\le \mbar,\qquad \mbar
={l \over 3\sqrt{3}}
\end{equation}
%
%
The black hole and cosmological horizon radii are plotted as a function of mass in Figure (\ref{horizonplot}).   As $m$ is increased towards its maximum value, the Nariai mass $\mbar$, the black hole and cosmological horizons converge, approaching the common value $r_b=r_c=l/\sqrt{3}$ for $m=\mbar$.

It is sometimes useful to parameterize SdS spacetimes in terms of the horizon radii $r_b$ and $r_c$, which are roots of $f(r)$, so that
\begin{equation}
f(r) = -{\Lambda\over 3r}(r-r_b)(r-r_c)(r-r_n)
\end{equation}
where $r_n =-(r_b+r_c)$ is the third, negative root.
Comparing with (\ref{SdS}), one obtains formulas for the mass and cosmological constant in terms of the horizon radii
\begin{equation}\label{mass}
m = {r_br_c(r_b+r_c)\over 2 (r_b^2+r_c^2+r_br_c)},\qquad l^2 = r_b^2+r_c^2+r_br_c
\end{equation}
%
%
%
The horizon temperatures $T_h = | f' (r_h ) |/4\pi $ are then given by
\begin{equation}\label{temperatures}
T_b={(r_c-r_b)(2r_b+r_c)\over 4\pi l^2\, r_b},\qquad
T_c={(r_c-r_b)(2r_c+r_b)\over 4\pi l^2\, r_c}
\end{equation}
%
These formulas show that both horizon temperatures approach zero when the horizon radii coincide,
in the limit $m=\mbar$. 

In our analysis below we will typically be fixing $\Lambda$, which is analogous to pressure, and consider varying the
size of the black hole. This is facilitated by yet another
parameterization of SdS black holes \cite{McInerney:2015xwa} in terms of deSitter length scale $l$ and the dimensionless quantity
\begin{equation}
\mu = {r_c-r_b\over l}
\end{equation}
which has the range $0\le \mu\le 1$.  In the Nariai limit ($r_b=r_c$) the parameter $\mu=0$, while in deSitter spacetime ($r_b=0$) one has $\mu=1$.  The horizon radii are given in terms of $(\mu,l)$ by
\begin{equation}\label{sdsrh}
r_b = {l\over 2}\left(-\mu +{2\over\sqrt{3}}\sqrt{1-{\mu^2\over 4}}\right),\qquad r_c = {l \over 2}\left(\mu +{2\over\sqrt{3}}\sqrt{1-{\mu^2\over 4}}\right)
\end{equation}
and the total entropy (\ref{totentropy}) has the simple form
\begin{equation}\label{sdsentropy}
S= {2\pi l^2\over 3} \left( 1+ {1\over 2} \mu^2 \right)
\end{equation}
The mass and horizon temperatures are given by
\begin{equation}\label{sdstemps}
T_b = {3(1-{\mu^2\over 2} +{\mu\over\sqrt{3}}\sqrt{1-{\mu^2\over 4}}\,\,)\mu\over 4\pi (1-\mu^2)l},\qquad 
T_c = {3(1-{\mu^2\over 2} -{\mu\over\sqrt{3}}\sqrt{1-{\mu^2\over 4}}\,\,)\mu\over 4\pi (1-\mu^2)l}
\end{equation}
\begin{equation}\label{sdsmass}
m = {(1-\mu^2)l \over 3\sqrt{3}}\sqrt{1-{\mu^2\over 4}}
\end{equation}

\begin{figure}[h]
\begin{center}\includegraphics[width=0.75\textwidth]{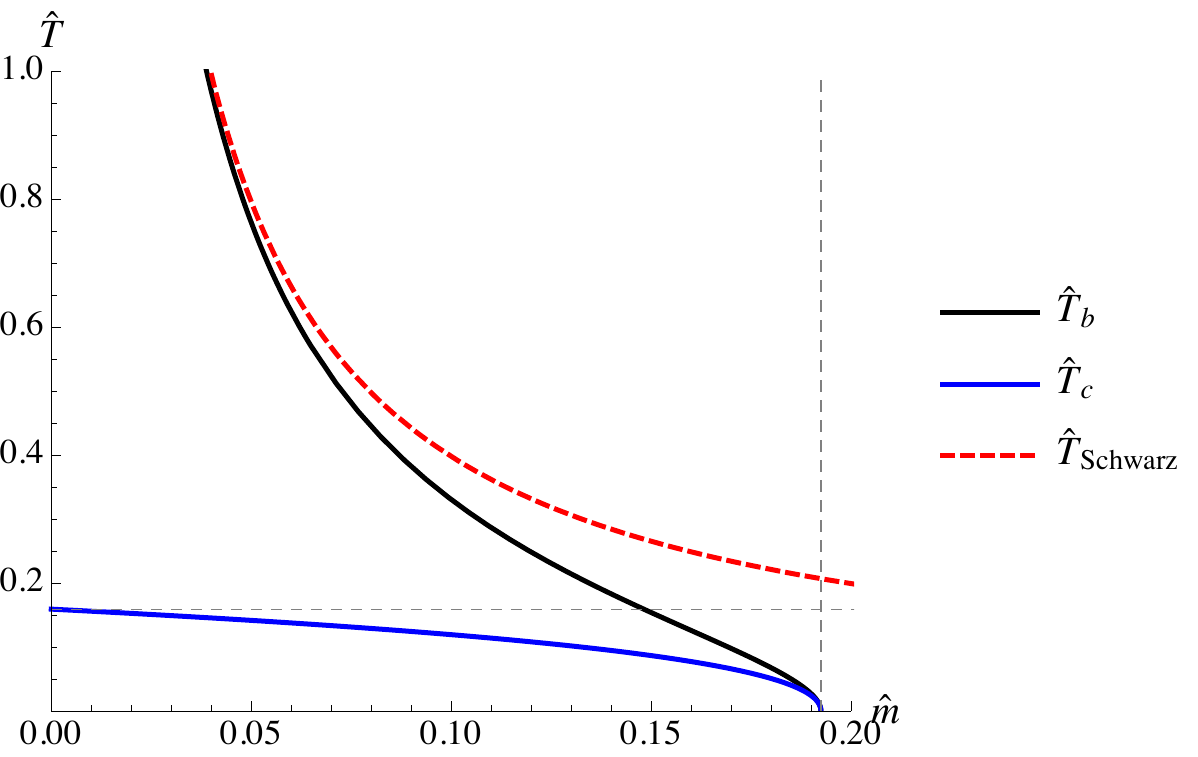}\end{center}
\caption{Plot of normalized black hole $\hat m \equiv m/l$ and cosmological horizon temperatures
 $\hat T_b$ and $\hat T_c$ versus the normalized mass $\hat m$.  The dashed horizontal line shows the deSitter temperature with no black hole $\hat T_c=1/2\pi\simeq 0.16$.  The vertical dashed line shows the maximum mass $\hat m = 1/ (3\sqrt{3})\simeq 0.19$.  The plot also shows normalized temperature for a Schwarzschild black hole $\hat T_{schwarz}= 1/ (8\pi \hat m)$ for comparison.} 
\label{temperaturefig}
\end{figure}

The parameter  $\mu$ is related to the thermodynamic volume $V$ \cite{Kastor:2009wy,Dolan:2013ft} which is conjugate to the cosmological constant in the extended phase space approach.  For SdS black holes, the thermodynamic volume of the causal patch between the black hole and cosmological horizons is equal to its Euclidean volume $V=4\pi (r_c^3-r_b^3)/3$ which can also be written as
\begin{equation}\label{vth}
V = {4\pi\over 3}\mu l^3
\end{equation}
Note that this thermodynamic volume can be measured by observers within the deSitter patch, in contrast to the mass $m$ which can be measured only at future infinity and is inaccessible to observers until then.
As a practical matter, the $\mu,\, l$ parameterization makes it straightforward to create various thermodynamic plots.  Figure (\ref{temperaturefig})  shows the black hole and cosmological horizon temperatures versus the mass.  We see that for small mass, the black hole temperature divergences, matching onto the Schwarzschild result, while the cosmological horizon temperature approaches its value in pure deSitter spacetime.  As the mass approaches its maximum value, both horizon temperatures approach zero.

\section{SdS black holes as constrained states}\label{bathsection}

Schwarzschild-deSitter spacetimes with mass in the range (\ref{range})  have both black and cosmological horizons, each with its own distinct temperature and entropy.  It is natural to ask how one should make sense of this in terms of a unified thermodynamic system.  Following \cite{Banks:2006rx}, we will think of the black hole as a constrained state of a deSitter heat bath.   We will see that the probability for such a state involves the total entropy $S=S_b+S_c$, which will be important in later sections of the paper.

The key to the heat bath interpretation is that deSitter black holes actually satisfy two first laws \cite{Dolan:2013ft}.  Restricting to fixed $\Lambda$ and vanishing angular momentum, the `first' first law gives the change in black hole entropy when the black hole mass is varied
\begin{equation}\label{firstfirst}
dS_b = {dm\over T_b}
\end{equation}
The new element for deSitter black holes is a `second' first law\footnote{A `third' first law is the sum of the `first' and `second'' first laws, $T_bdS_b+T_cdS_c=0$, which eliminates the black hole mass and refers only to quantities at the boundaries of the deSitter causal patch.} giving the change in cosmological horizon entropy with a change in black hole mass
\begin{equation}\label{secondfirst}
dS_c = - {dm\over T_c}
\end{equation}
We interpret the opposing signs in these two statements as indicating that as energy is added to the black hole, it is drawn from a deSitter bath of energy, reducing the entropy of the bath.

Recall how Boltzmann statistics arise in the canonical ensemble.  One has a heat bath at fixed temperature $T_c$ with entropy $S_c(E)$ where $E$ is the energy in the bath.  The energies of microstates of a subsystem in thermal contact with the bath are given by $m$.    Assume to start that these microstates are non-degenerate.  Take the total energy of the bath plus the system to be $\mathbb E$.  If we assume that all microstates of the combined system of subsystem plus bath are equally probable, then the probability of the system being in a microstate with energy $m$ is proportional to the number of microstates of the bath with energy $E={\mathbb E}-m$
\begin{equation}\label{basicprob}
P(m) \propto e^{S_c({\mathbb E}-m)}
\end{equation}
Assuming that $m\ll E$, one can approximate the entropy of the bath by
\begin{align}
S_c({\mathbb E}-m) & \simeq S_c(\mathbb E) - ({\partial S_c\over \partial E})\, m\\
&\simeq S_c(\mathbb E) - {m\over T_c}
\end{align}
Since the first term is independent of $m$, this can be included in the normalization factor, and we obtain the standard Boltzmann factor for the probabilities 
\begin{equation}
P(m) \propto e^{-m/T_c}
\end{equation}
If there are some number of degenerate microstates of the subsystem having energy $m$, then the probability (\ref{basicprob}) is multiplied by the number of such microstates, which will be the exponential of the entropy of the subsystem, which we will denote by $S_b(m)$.  We then have
\begin{equation}
P(m)\propto e^{S(m)}
\end{equation}
where the $S(m)$ is the total entropy of the bath and subsystem $S(m)=S_b(m)+S_c({\mathbb E}-m)$.  In the non-degenerate case, the subsystem also naturally has a temperature $T_b$ associated with it and given by
\begin{equation}
{1\over T_b} = {\partial S_b\over \partial m}
\end{equation}
Identifying the deSitter horizon with the bath and the black hole with the subsystem provides a unified thermodynamic interpretation, which accounts for the independent temperatures of the deSitter and black hole horizons, and gives the combined entropy of the black hole and deSitter horizons a clear physical significance.  The black hole in this picture is a non-equilibrium `constrained state' of the overall system, in which energy $m$ is contained in the subsystem.

\section{Schottky anomaly of SdS black holes}\label{twolevelsec}

\begin{figure}
\centering
\begin{subfigure}{.5\textwidth}
  \centering
  \includegraphics[width=.9\linewidth]{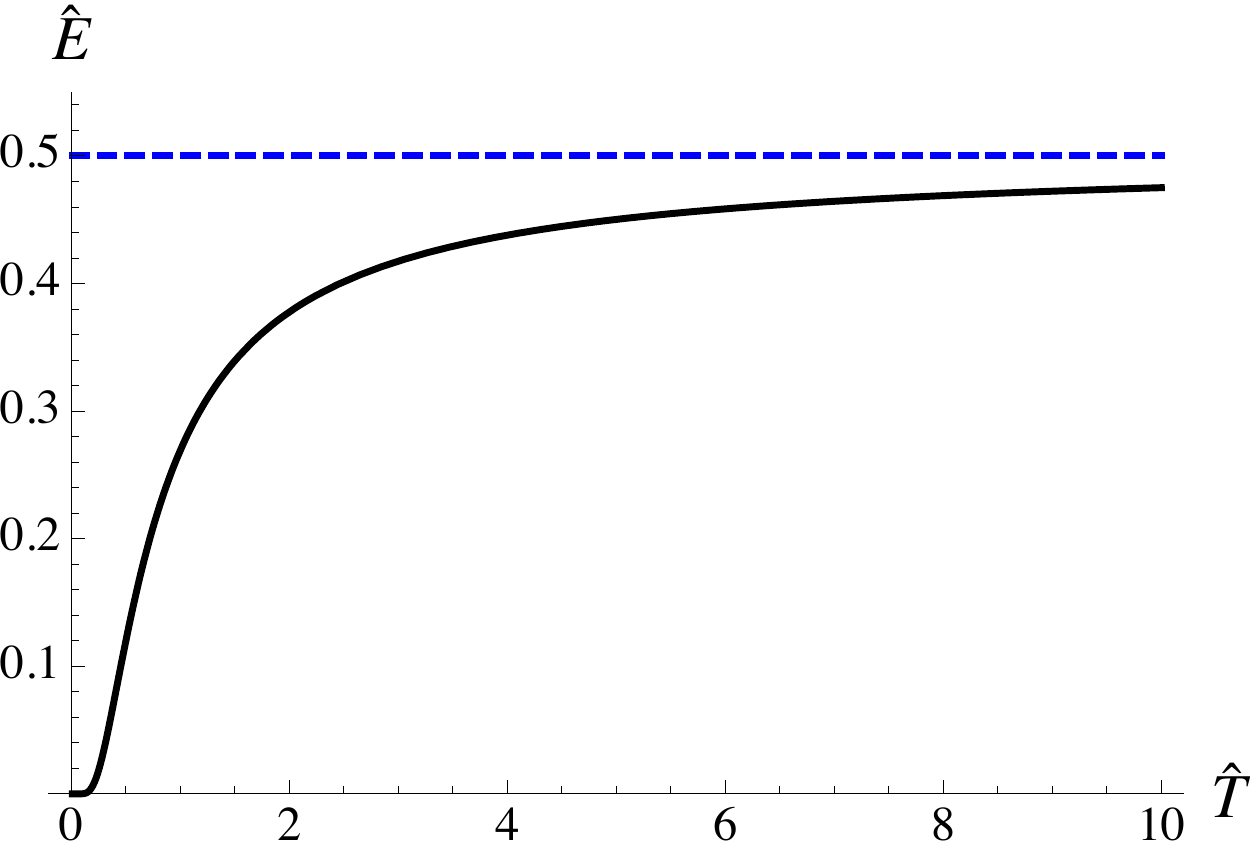}
  \label{fig:sub1}
\end{subfigure}%
\begin{subfigure}{.5\textwidth}
  \centering
  \includegraphics[width=.9\linewidth]{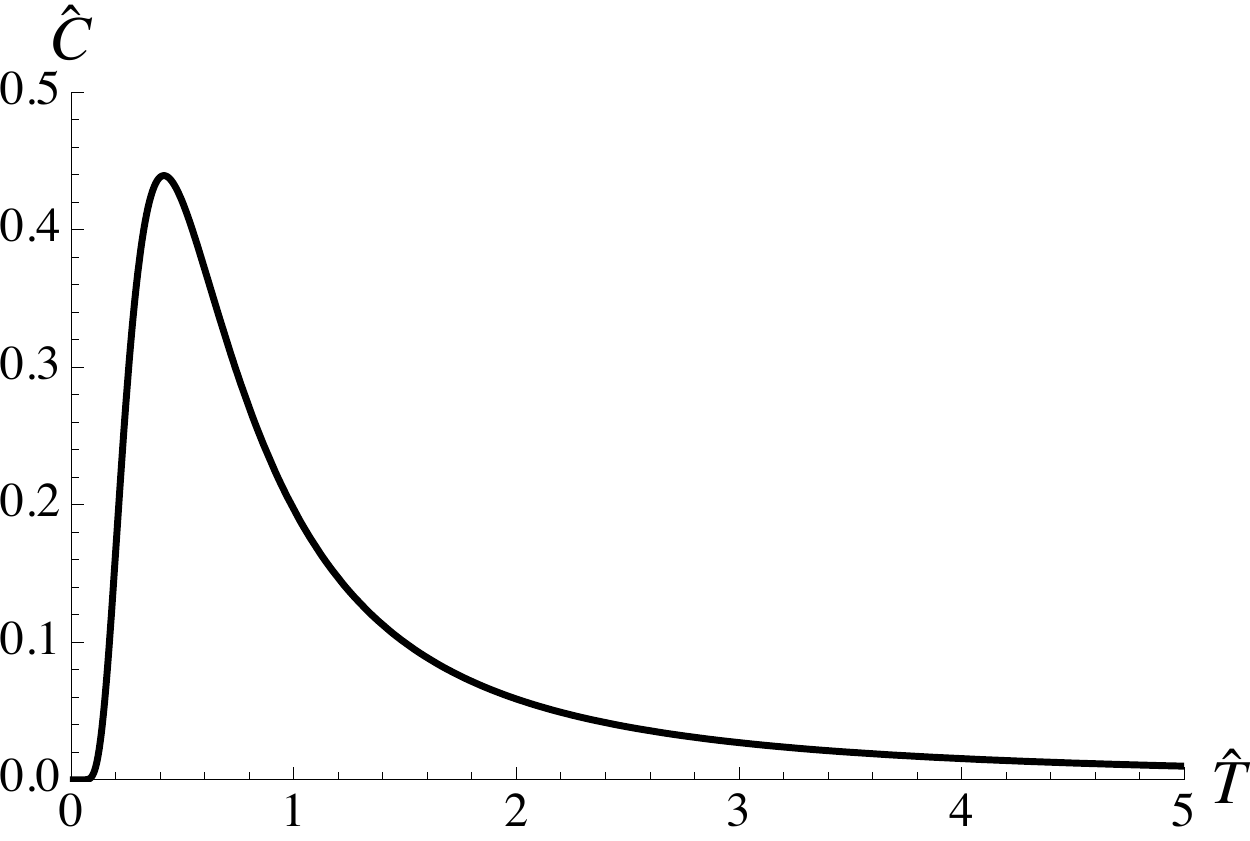}
  \label{fig:sub2}
\end{subfigure}
\caption{Plots of the normalized energy $\hat E \equiv E/ \varep$ and heat capacity $\hat C\equiv C/\varep^2$ versus normalized temperature $\hat T\equiv T/\varep$ for the two-level system.  The high temperature limits of the normalized energy and entropy are $\hat E=1/2$ and $S=\ln 2$.}
\label{twolevelenergy}
\end{figure}

The heat capacity of a thermal system is typically an increasing function of temperature.  However, a peak in the heat capacity, known as a Shottky anomaly can occur in a system that has a maximum energy.  Before discussing this in the context of SdS black holes, consider the  
simple example of a two level system having vanishing ground state energy and a single excited state with energy $\varep$.
Working in the canonical ensemble, 
%
%
the average energy of the system as a function of temperature is then given by
\begin{equation}
E (T)= {\varep e^{-\varep/T}\over 1 + e^{-\varep/T}}
\end{equation}
%
and is plotted on the left in Figure (\ref{twolevelenergy}).  The average energy is a monotonically rising function of temperature, approaching the limiting value of $E=\varep/2$ at high temperatures where the two states have essentially equal probabilities.  The entropy is given by
\begin{equation}
S(T) = \log(1+e^{-\varep/T})+{e^{-\varep/T}\over T(1+e^{-\varep/T})}
\end{equation}
and its behavior is qualitatively the same as the energy.  The entropy vanishes at low temperatures, where only the ground state has significant probability and asymptotes to $\log 2$ in the limit of high temperature as both states become equally probable.


Now consider the heat capacity, $C\equiv dE/dT$, of the two level system which is given by
\begin{equation}
C= {\varep^2\over T^2}{e^{-\varep/T}\over (1 + e^{-\varep/T})^2}
\end{equation}
and shown on the right in Figure (\ref{twolevelenergy}).   We see that the heat capacity has a prominent peak, occurring at $T\simeq 0.417\, \varep$, while approaching zero in the limits of both high and low temperatures.  This behavior is easily understood in physical terms.
At very low temperatures, with $T\ll\varep$, a small increase in temperature is insufficient to appreciably change the occupation of the excited state and hence leaves the average energy almost unchanged.  This corresponds to the flat behavior of the energy in Figure (\ref{twolevelenergy}) for very low temperatures.
At high temperatures, with $T\gg\varep$, the average energy approaches its maximum value $E=\varep/2$ and raising the temperature further again results in very little change. The temperature scale at which the transition happens is
set by the energy gap $\epsilon$.  Note that the Schottky anomaly will also be reflected in the quantity
\begin{equation}
{\partial S\over \partial T} = {C\over T}.
\end{equation}
%



\begin{figure}
\centering
\begin{subfigure}{.5\textwidth}
  \centering
   \includegraphics[width=.9\linewidth]{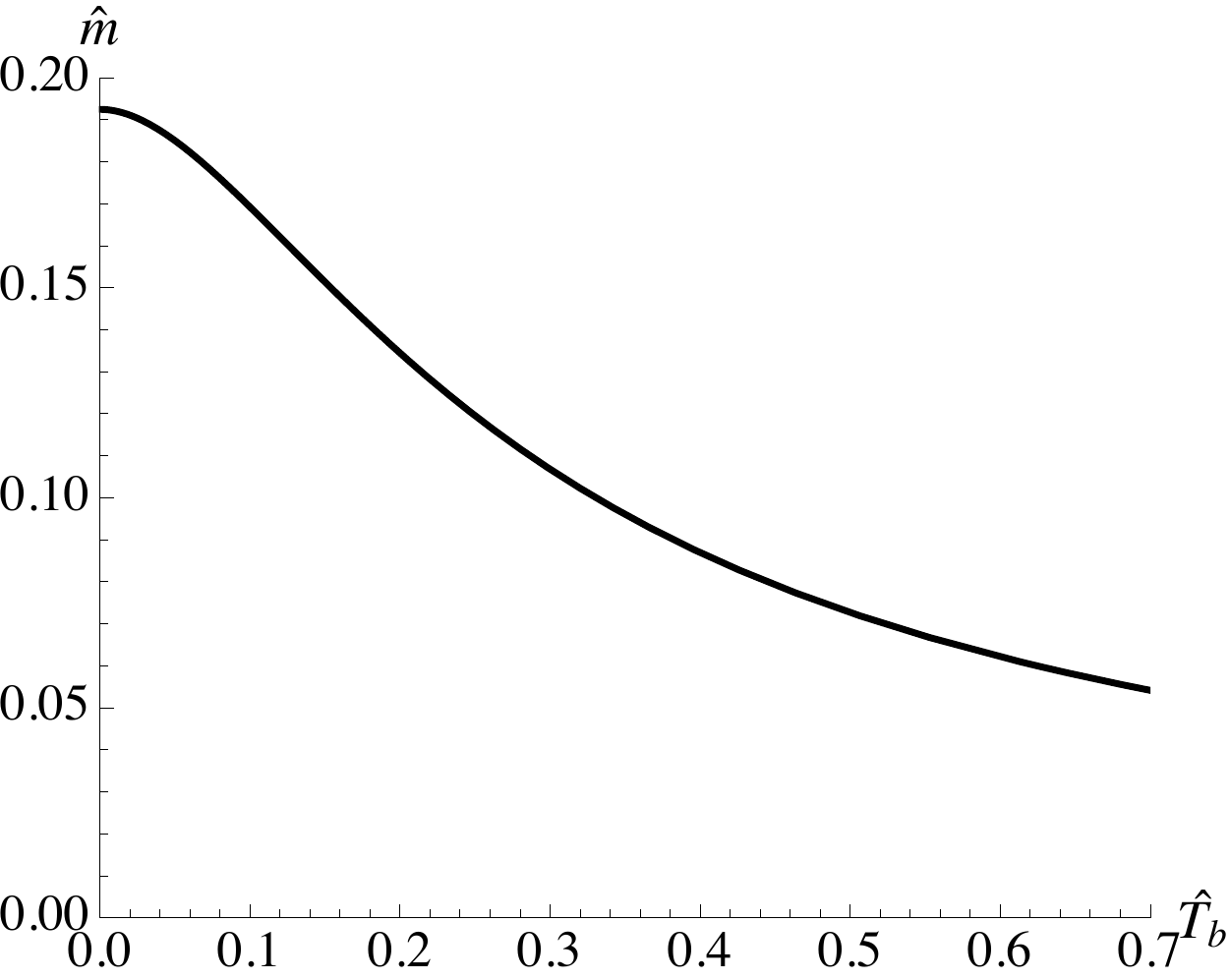}
  \label{fig:sub1}
\end{subfigure}%
\begin{subfigure}{.5\textwidth}
  \centering
  \includegraphics[width=.9\linewidth]{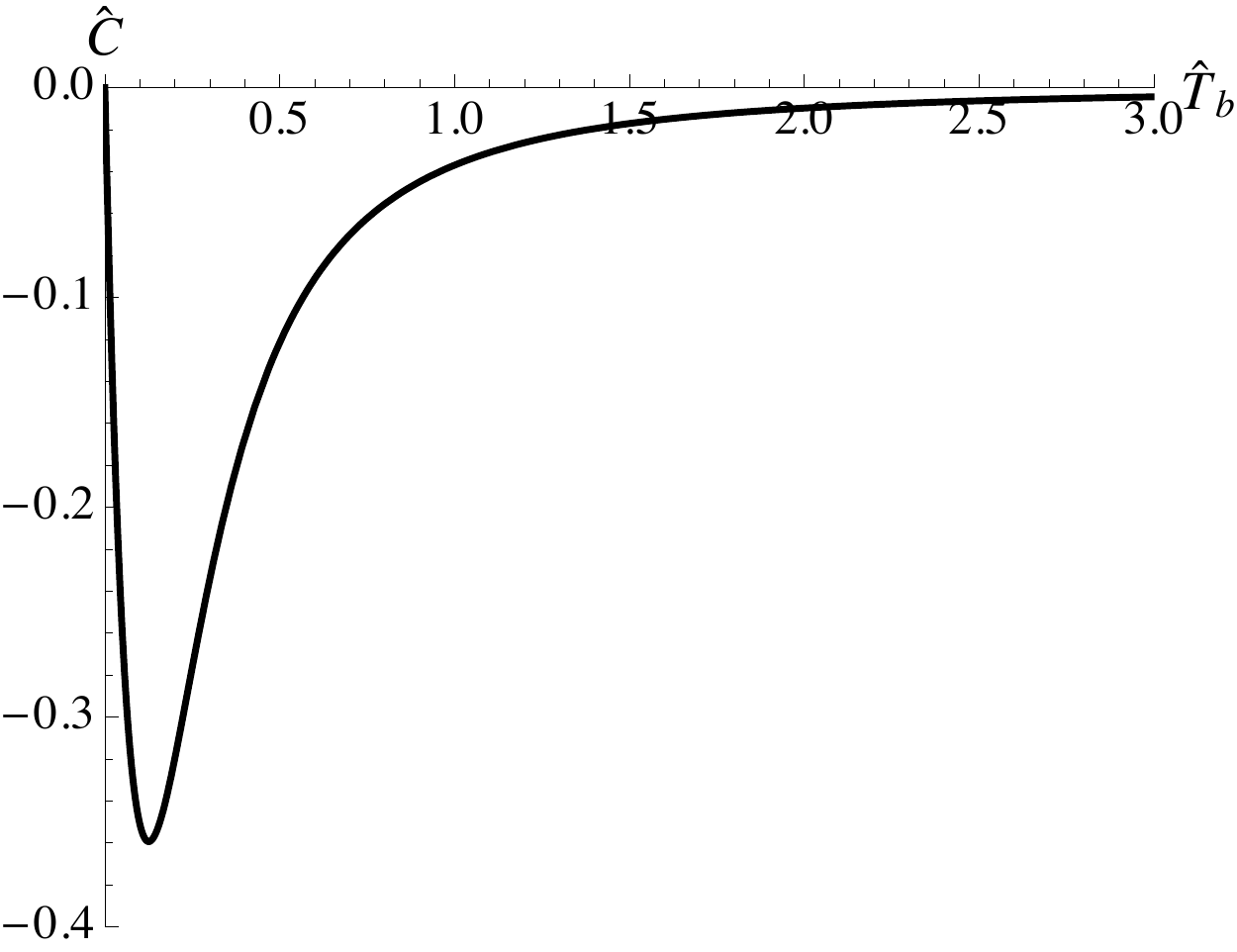}
  \label{fig:sub2}
\end{subfigure}
\caption{Plot of normalized black hole mass $\hat m\equiv m/l_c$ and normalized heat capacity versus normalized black hole temperature $\hat T_b\equiv l \,T_b $.}
\label{massvtemp}
\end{figure}

We now return to Schwarzschild-deSitter black holes.  If we focus on the black hole mass as a measure of the energy of the system, then there is a clear resemblance to the two state system described above.  The energy runs over a finite range, with the black hole mass satisfying $0\le m\le \mbar$, while the black hole horizon temperature runs over an infinite range.  Of course, the association of temperatures is reversed in the black hole case.  The temperature diverges in the small black hole limit and goes to zero for the Nariai mass $\mbar\simeq 0.19\, l$.  Figure (\ref{massvtemp}) shows the mass versus black hole temperature on the left and the black hole heat capacity $C$ on the right.  We see that the heat capacity has an inverted Schottky peak, corresponding to the inflection point in the plot of black hole mass.


%

\begin{figure}
\centering
\begin{subfigure}{.5\textwidth}
  \centering
  \includegraphics[width=.9\linewidth]{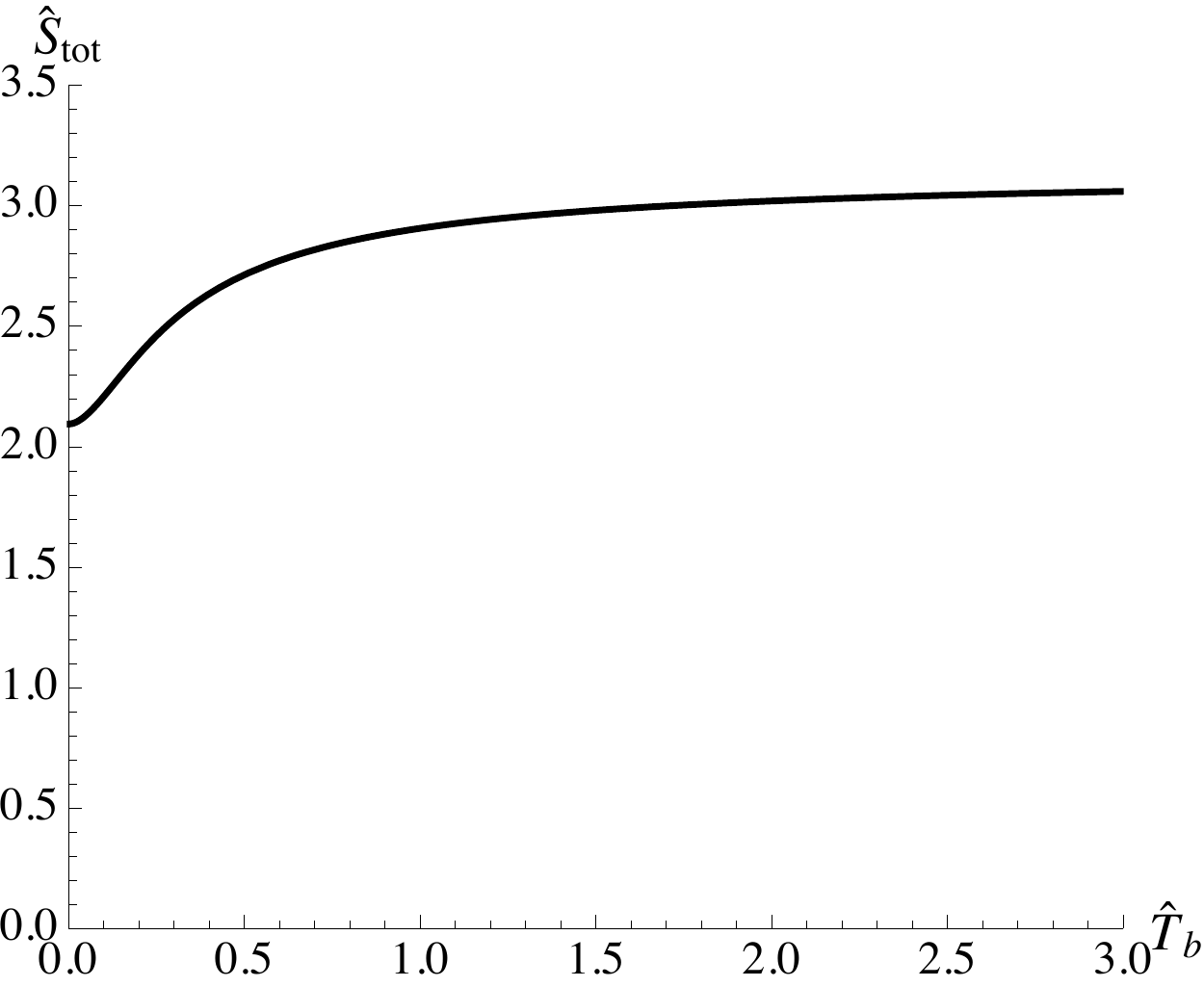}
  \label{fig:sub1}
\end{subfigure}%
\begin{subfigure}{.5\textwidth}
  \centering
  \includegraphics[width=.9\linewidth]{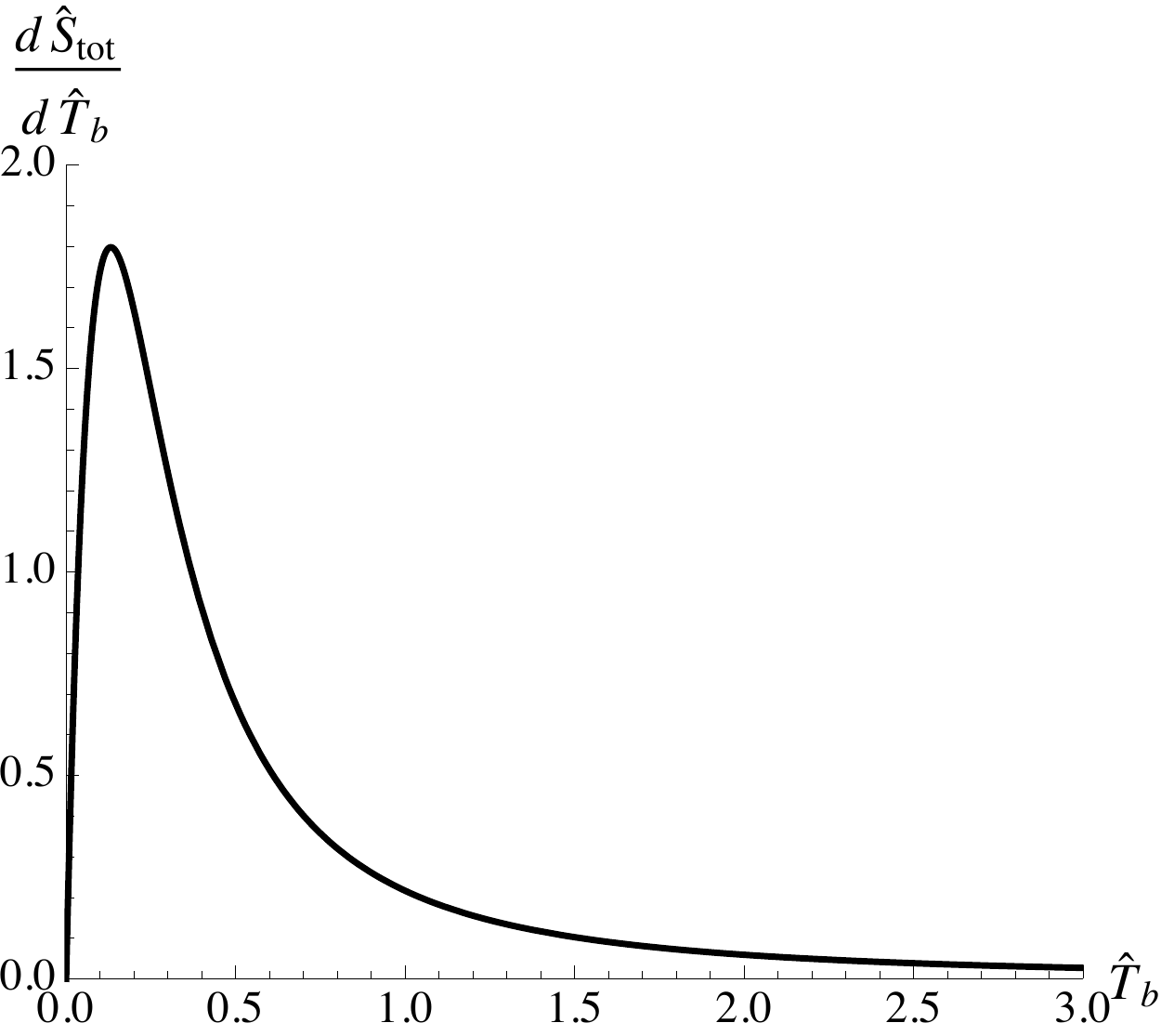}
  \label{fig:sub2}
\end{subfigure}
\caption{Left: normalized black hole mass $\hat m\equiv m/l$ and normalized total entropy versus normalized black hole temperature $\hat T_b\equiv l\,T_b $. Right: $dS_{tot}/dT_b$ exhibits a peak.}
\label{entropyfig}
\end{figure}

Similar Schottky type behavior can be seen in the dependence of the total entropy 
$S=S_b +S_c$ of the black hole and cosmological horizons as a function of black hole horizon temperature.  We argued in the last section that the total entropy is a thermodynamically important quantity.  The total entropy is a decreasing function of black hole mass and hence an increasing function of black hole horizon temperature, as shown in the left hand plot of Figure (\ref{entropyfig}).  The curve also has an inflection point and hence its derivative, shown on the right, has a peaked behavior,  in this case without an inversion.


We can also examine the plateau type behavior for SdS black holes that leads to the Schottky anomaly analytically.
Let's start by examining the behavior of the total entropy $S=S_b+S_c$ in the limits of small and large black hole mass.  We will make use of a linear combination of the two first laws (\ref{firstfirst}) and (\ref{secondfirst})
\begin{equation}\label{causalfirst}
T_b dS_b+T_cdS_c =0
\end{equation}
As discussed in \cite{Dolan:2013ft} this relation emerged from carrying out the first law construction in the causal patch between the black hole and deSitter horizons, so that the black hole mass which comes from a boundary integral at infinity does not contribute.  Eq.~(\ref{causalfirst}) implies that the variation in the total entropy is related to a variation in the black hole entropy by
\begin{equation}\label{varent}
dS = -\left ({T_b\over T_c}-1\right ) dS_b
\end{equation}
Since the black hole temperature is always greater than or equal to the deSitter temperature, one immediately sees that the total entropy always decreases if the size of the black hole is increased, yielding the monotonically increasing behavior shown on the left in Figure (\ref{entropyfig}).


Now consider the small black hole limit such that $T_b\gg T_c$.  In this limit the thermodynamic properties of the black hole are approximately that of Schwarzschild.  In particular, we have $T_b\simeq 1/(8\pi r_b)$, which leads to the relation
\begin{equation}\label{bhent}
dS_b \simeq -{dT_b\over 8\pi T_b^3 }
\end{equation}
%
In this limit the cosmological horizon temperature is approximately equal to its value
in pure deSitter, $T_c \simeq 1/(2\pi l_c )$.   Combining these ingredients, we find in this limit that
\begin{equation}\label{stotsmall}
dS\simeq {l\over 4 T_b^2} dT_b
\end{equation}
which vanishes as $T_b\rightarrow\infty$, corresponding to the flattening behavior of the entropy curve on the left hand side of Figure (\ref{entropyfig}) in the high temperature limit.


The zero temperature limit for the black hole occurs when the black hole mass approaches the maximal mass $m_N$ and the black hole and cosmological horizons approach one another.  Approaching this limit, the ratio of temperatures in (\ref{varent}) becomes $T_b/T_c\simeq 1$, giving $dS\simeq 0$ and thus another plateau in the entropy at low temperatures, as seen on the left of Figure (\ref{entropyfig}).  Together, the plateaus in the entropy at large and small temperatures account for the Schottky-type peak occurring in the plot of $dS/dT_b$ on the right hand side of Figure (\ref{entropyfig}).


Returning to the heat capacity, the  black hole mass shown on the left hand side of Fig.~(\ref{massvtemp}) also flattens in the limits of high and low temperature, giving rise to the (inverted) Schottky peak in the heat capacity.  Let us see how this behavior arises analytically starting from the `first' first law (\ref{firstfirst}). Combining this in the high temperature limit with Eq.~(\ref{bhent}),  we find
\begin{equation}
dm = -{dT_b\over 8\pi T_b^2}
\end{equation}
which describes  the flattening at high temperature.  At low temperature, the first law (\ref{firstfirst}) implies a plateau so long as the rate of change $dS_b/dT_b$ remains finite.  To establish that this is the case, we can write
\begin{equation}
{dS_b\over dT_b} = \left({dT_b\over d\mu}\right)^{-1}  {dS_b\over d\mu}
\end{equation}
and use equations (\ref{sdsrh})-(\ref{sdstemps}) to show that in the low temperature limit ($\mu\rightarrow 0$) 
\begin{equation}
{dT_b\over d\mu}\simeq {3\over 4l},\qquad {dS_b\over d\mu}\simeq -{\pi l^2\over \sqrt{3}}
\end{equation}
which are indeed finite.  The first law then implies that the heat capacity, shown on the right hand side of Figure (\ref{massvtemp}), vanishes linearly for low temperatures.

In this section, we examined two indicators of Schottky-type peaked behavior in the thermodynamics of Schwarzschild-deSitter black holes. 
Both involved the black hole horizon temperature, which runs over an infinite range.  However, both also depended crucially on the deSitter horizon.  In the case of the heat capacity $dm/dT_b$, the black hole mass has the finite upper limit $m_N$ due to the presence of the deSitter horizon, while in the case of $dS/dT_b$ the total entropy includes that of the deSitter horizon as well as the black hole entropy.

\section{Spin systems}\label{spinsection}

We have seen that the Schottky anomaly arises naturally in the thermodynamics of SdS black holes, reflecting the joint influences of the black hole and cosmological horizons.  It is interesting to try to model the underlying quantum SdS degrees of freedom by a simple paramagnetic spin system which exhibits similar behavior.  As discussed in Sec. (\ref{bathsection}), SdS black holes can be thought of as constrained states of pure dS. Accordingly, our spin model should include a large number of degrees of freedom representing a deSitter `heat bath' and a subsystem with a smaller number of degrees of freedom representing the black hole.  The simple two-level system  described at the beginning of Section (\ref{twolevelsec}) is  insufficient for this purpose, but we can generalize to a large number $N$ of non-interacting spins, each of which have two possible energy levels. In this section we will see which aspects of SdS thermodynamics are reflected by this model, as well as where they diverge.

\begin{figure}
\centering
\begin{subfigure}{.5\textwidth}
  \centering
    \includegraphics[width=.9\linewidth]{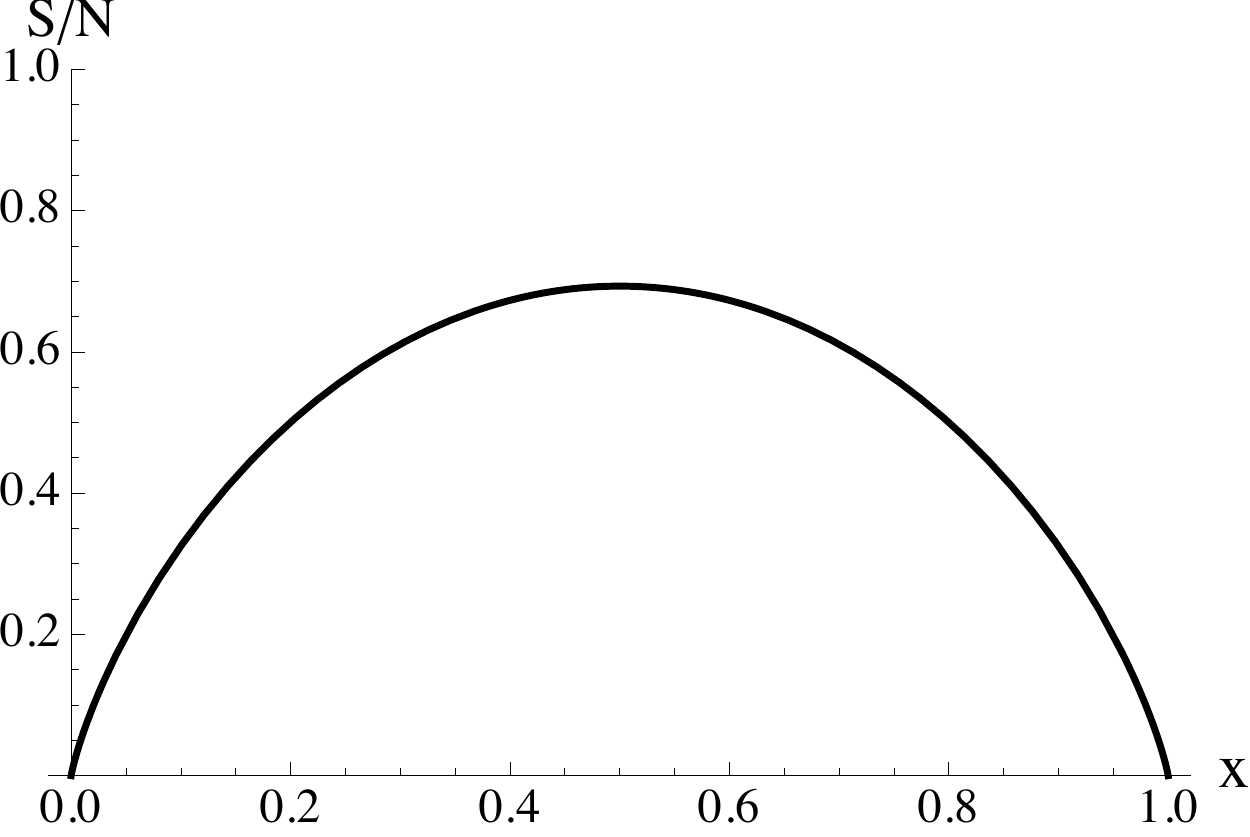}
\end{subfigure}%
\begin{subfigure}{.5\textwidth}
  \centering
    \includegraphics[width=.9\linewidth]{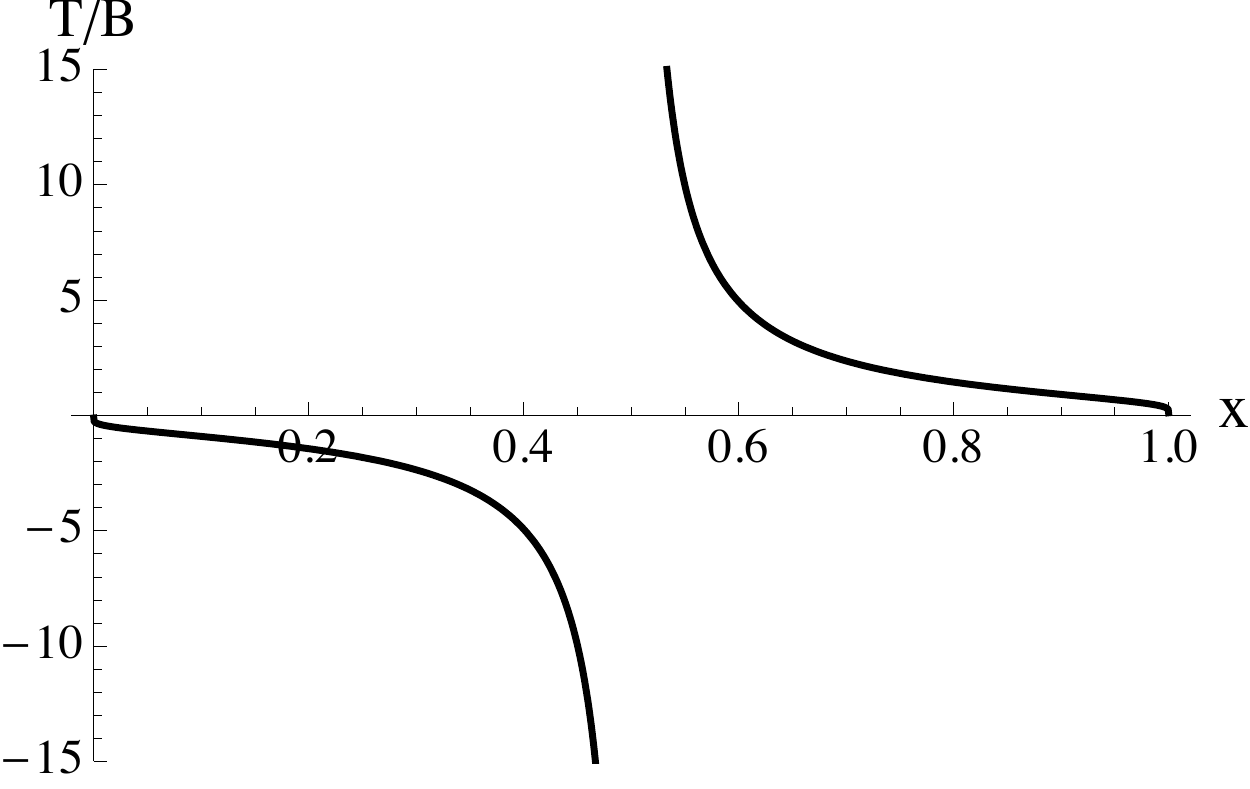}
\end{subfigure}
\caption{Entropy per spin (left) and temperature ($T/B$, right) plotted as a function of $x=N_+/N$.}
\label{spinfig}
\end{figure}

Consider a system of $N$ spins of unit magnetic moment interacting with a magnetic field $B$.  The energy of a configuration of spins is 
\begin{equation}
 U= B (N_- -N_+)= B(N-2N_+)
\end{equation}
where $N_\pm$ count the number of spins parallel and anti-parallel to the magnetic field.  We consider the case that $N\gg 1$ and work in the microcanonical ensemble. Corresponding results in the canonical ensemble will be equivalent up to fluctuations that scale as $1/\sqrt{N}$.  The number of microstates with fixed energy is  given by the binomial coefficient
\begin{equation}
\Omega(N_+) = {N\choose N_+} = {N!\over N_+!(N-N_+)!}
\end{equation}
and for large $N$ the entropy per spin is approximately
\begin{equation}
S/N\simeq -x\log x -(1-x)\log (1-x)
\end{equation}
where $x\equiv N_+/N$.  The entropy vanishes for $x=0$ and $1$ and is maximized for $x=1/2$ as shown on the left in Figure (\ref{spinfig}).
The temperature is given by
\begin{equation}\label{spintempformula}
{1\over T}={1\over 2B}\log\left({x\over 1-x}\right)
\end{equation}
and is shown on the right in Figure (\ref{spinfig}).   As one would expect, all spins are aligned with the magnetic field ($x=1$) in the low temperature limit, while in the high temperature limit aligned and anti-aligned spins occur in equal numbers ($x=1/2$).  The temperature is negative over the range $0<x<1/2$ in which the higher energy anti-aligned state is more populated, reflecting the thermodynamic instability of such states.

In order to model an SdS black hole with this simple  spin system, we want to pick a fraction $x$ of aligned spins to represent the state of the whole system, and also a number of spins $N_b<N$ with fraction $x_b$ of aligned spins to represent the black hole subsystem.  

The whole system, with no subsystem specified, represents empty deSitter spacetime.  The fraction $x$ of aligned spins  determines the deSitter temperature through Eq.~(\ref{spintempformula}). 
If we now add a small black hole, then its temperature (given by (\ref{spintempformula}) with $x\rightarrow x_b$) should be large, so we should pick $x_b$ to be just slightly greater than $1/2$.  In particular, we will want $x_b<x$ so that the black hole is hotter than the unperturbed deSitter horizon. Presumably we should also require $N_b\ll N$.  Let us see what happens to the deSitter temperature in the presence of the small black hole.

The cosmological horizon is now represented by the remaining $N_c=N-N_b$ spins which have an aligned fraction
\begin{equation}
x_c = {Nx-N_bx_b\over N-N_b}.
\end{equation}
In the limit that $N_b\ll N$, this is approximately given by
\begin{equation}
x_c\simeq x +(x-x_b){N_b\over N}.
\end{equation}
Since the small black hole is hotter than the original deSitter temperature ($x_b<x$), we see that the cosmological horizon temperature is reduced by the presence of the back hole, in agreement with the behavior  shown in Figure (\ref{temperaturefig}).  This will continue to be the case for any value of $N_b<N$, so long as $x_b<x$.  Segregating off more anti-aligned spins in the black hole subsystem increases the concentration of aligned spins in the remaining cosmological bath.

Increasing the size of the black hole should correspond to increasing $N_b$, while also increasing $x_b$ so that the black hole temperature decreases. This leads us to note a number of features that work less well in this model.  First, the energy in the black hole subsystem is given by
\begin{equation}
U_b = BN_b(1-2x_b)
\end{equation}
which is negative, so long as we stay in the regime ${1\over 2}< x_b\le 1$ with positive temperature.  Moreover, if we increase the size of the black hole by increasing both $N_b$ and $x_b$, this energy becomes more negative.  A second feature that is partly but not quite right concerns the large black hole limit. If $x_b$ is increased all the way to $x$ then the black hole and cosmological temperatures will coincide.  However, the common horizon temperature will be equal to the original deSitter temperature of the system, rather than vanishing as it does in the Nariai limit.

\begin{figure}[t!]
\begin{center}
\includegraphics[width=0.5\linewidth]{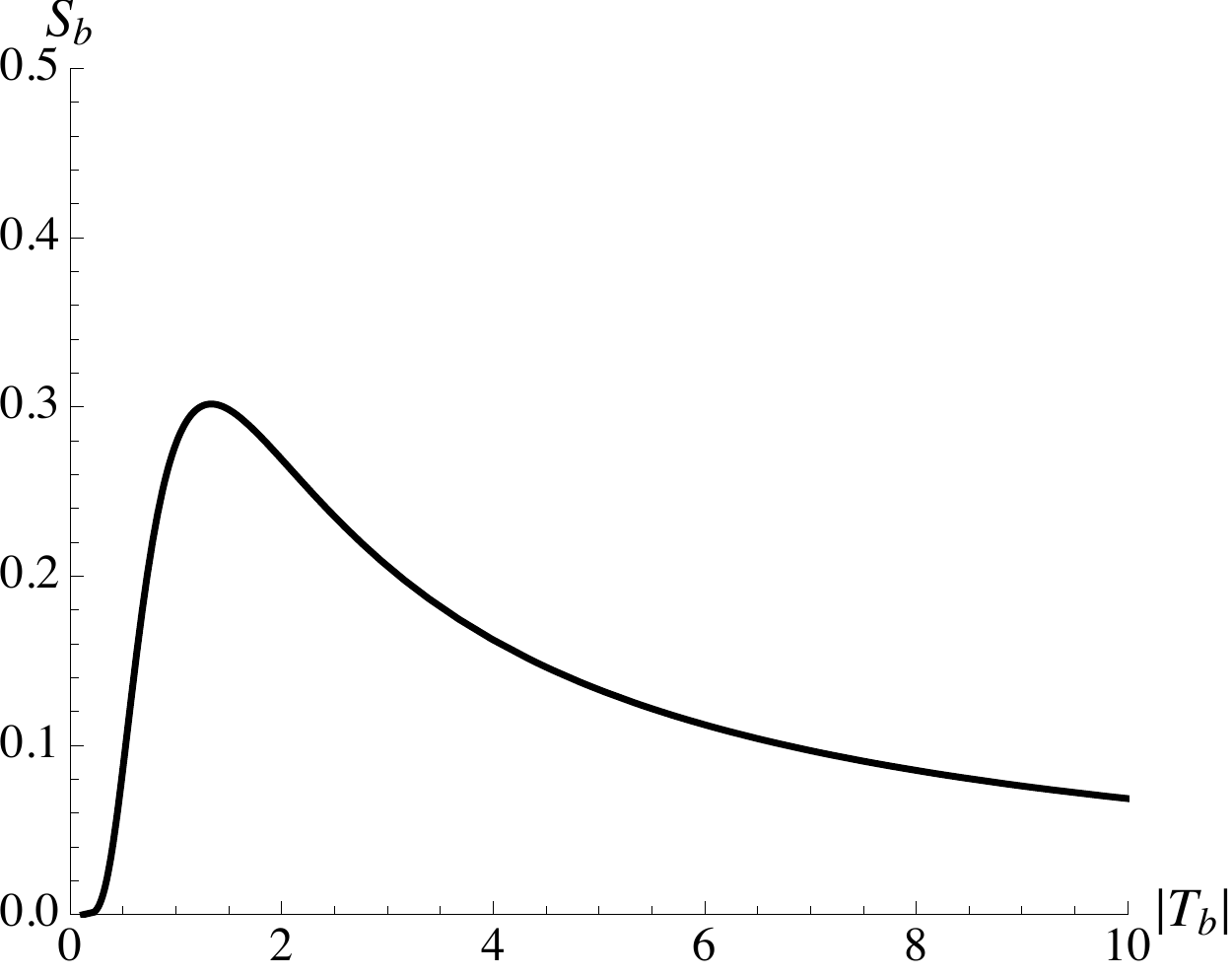}~~~~~~
\includegraphics[width=0.5\linewidth]{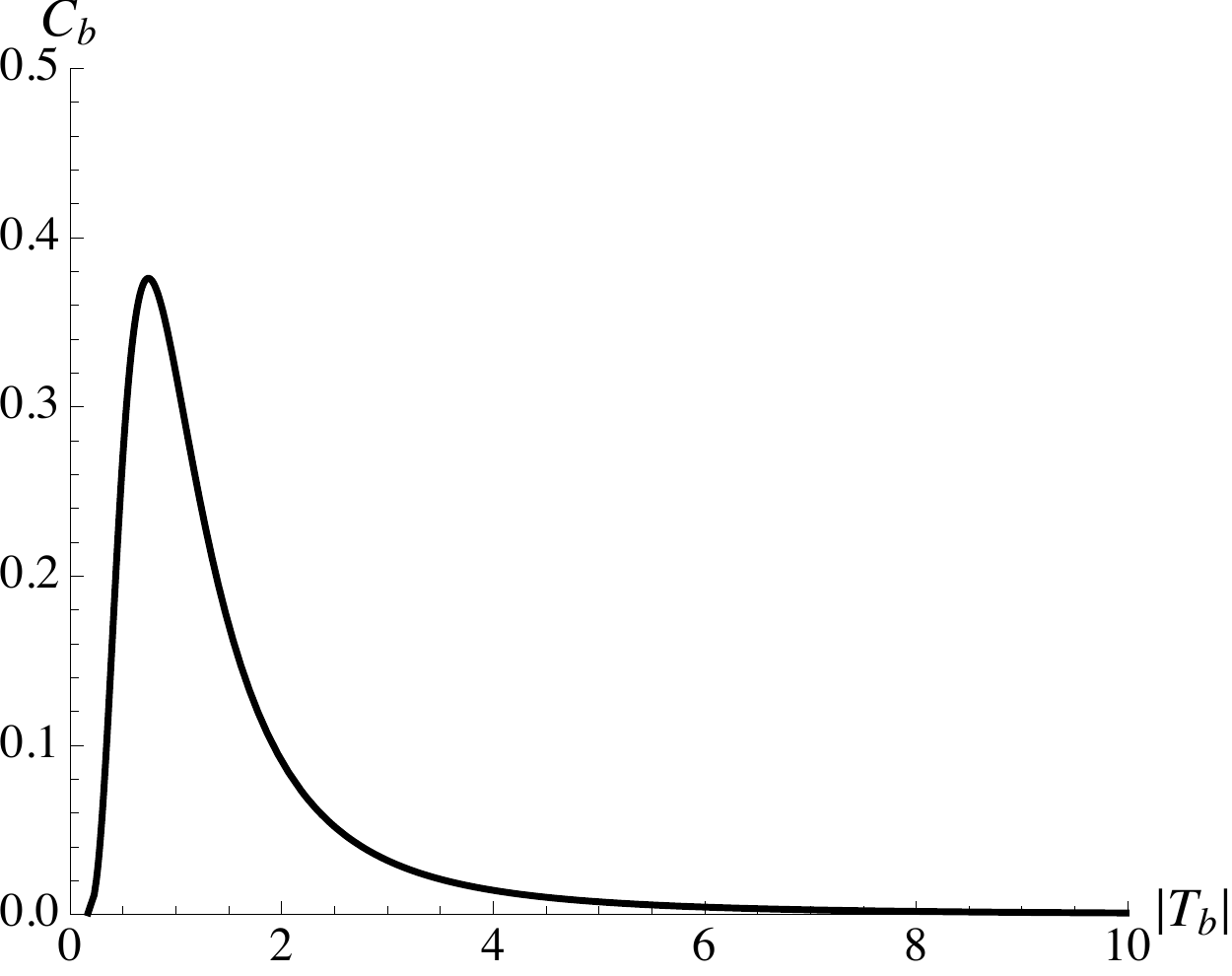}
\caption{Left: the entropy of the ``black hole" in the spin model defined by Eq.~(\ref{eq:spinbh2}), normalized to $N_-$, the total number of down spins. Right: The heat capacity exhibits the Schottky anomaly.} 
\label{fig:spinSbTb}
\end{center}
\end{figure}

We can partially address these differences by considering a refinement of the model above. First, we take  $0< x_b< {1\over 2}$. This restriction makes the energy of the black hole positive (although its temperature will be negative). Second, we can adjust the family of constrained states, prescribing an $N_b(x_b)$ so that as $x_b\rightarrow 0$ the black hole has absorbed all of the anti-aligned spins. For example, we can take
\begin{align}
N_b(x_b)=(1-2x_b)(1-x)N,
\label{eq:spinbh2}
\end{align}
which attains its largest value, $N_-$, as $x_b\rightarrow 0$. Therefore $x_b\rightarrow 0$ is the large black hole limit. The entropy of the subsystem is shown in Fig.~\ref{fig:spinSbTb}. In this model the temperature is unbounded and the entropy is bounded, as desired, and the temperature of both the black hole and the cosmological horizon go to zero in the large black hole limit. However, we have not completely ameliorated the differences between the model and SdS: for finite-size black holes the temperature is negative. Also, for small $|T_b|$, the entropy goes to zero. 


The model does reflect the Schottky anomaly, as expected. The heat capacity is shown in Fig.~\ref{fig:spinSbTb} and exhibits the Schottky peak, although the sign is opposite that of the true SdS case.

\section{Quantum fluctuations}\label{quantumsec}

The Schottky anomaly of SdS black holes is associated with the plateaus found in plots of the mass and total entropy in the limits of high and low black hole horizon temperature.  It is natural to ask if these features persist in the presence of quantum fluctuations, which are an integral part of black hole thermodynamics.   
Making use of recent calculations of black hole and cosmological particle production in SdS \cite{qiu}, we will see 
that the quantum spectra, total number of produced particles, and the energy and entropy in those particles, indeed show
evidence of saturation at high and low black hole temperature.


Hawking \cite{Hawking:1974sw} originally calculated that asymptotically flat black holes emit a flux of quantum particles with a black body
spectrum at temperature $T_b = \kappa_b/2\pi$, where $\kappa_b$ is the surface gravity of the black hole horizon.
Gibbons and Hawking \cite{Gibbons:1977mu} subsequently 
showed that the cosmological horizon in deSitter is also a source of quantum emission, finding a thermal 
spectrum with  $ T_c = |\kappa_c|/2\pi$, where $\kappa_c$ is now the (negative) surface gravity of the cosmological horizon. 
In black hole spacetimes with $\Lambda >0$ there is particle production due to both the black hole and 
cosmological horizons. Since in general
the black hole and cosmological horizon temperatures will not be the same, the nature of the thermodynamic system described by deSitter black holes has remained somewhat unclear.
However, as discussed in Section (\ref{spinsection}) a two-temperature characterization naturally
arises for constrained states of a statistical system. 
Thermal properties of SdS black holes were discussed in \cite{Gibbons:1977mu} based on a variety of physical arguments, 
while a calculation of particle production for RNdS
black holes was carried out in \cite{Kastor:1993mj} focusing on the 
the  $|Q| =M$ case where the black hole and cosmological horizon temperatures are equal and nonzero.
It was found in \cite{Kastor:1993mj} that the particle spectra are not thermal, a feature shared by 
the spectra presented  below.

More recently \cite{Bhattacharya:2018ltm} has computed the black hole and cosmological particle production in SdS spacetimes, choosing a set of particle states such that both spectra are thermal at temperatures respectively of $T_b$ and $T_c$. However, in this formulation the particle states are not well behaved on the horizons. 
In the present analysis we will be particularly interested in fluctuations crossing the horizons, so we require modes
that are well behaved there. 
Following \cite{Kastor:1993mj}, Ref.~\cite{qiu} computes particle production in terms of states defined using local Kruskal coordinates. 
Let us summarize
the results of \cite{qiu} that are useful for consideration of the Schottky anomaly.

Consider the causal diamond region of an SdS spacetime bounded by the past and future black hole and cosmological horizons.  Compared to an asymptotically flat black hole, the cosmological horizons have replaced the boundaries at past and future null infinity.   We will interpret particles that cross the future cosmological horizon as having been produced by the black hole, and those entering the black hole horizon as coming from the cosmological horizon. 
The expected number of produced particles with frequency $\omega$ that 
are absorbed per unit time by the horizon $h$, where $h=b,c$ denote the black hole and cosmological horizons, is given by the product of the density of states $\rho (\omega )  d\omega =   R\sum_l (2l+1) d\omega$  times the quantum spectrum, and has the form
  \bea\label{spectraldens}
\N^h  & = & \langle N_\omega^h \rangle \rho (\omega )\, d\omega   \\
&=&  {\omega^2 r_b^2\over R} \, \left(\int d\omega' |\beta^h_{\omega  \omega' l } |^2\right)\,  d\omega \ , \quad \omega >\omega_0 
\eea
 where $\beta^h_{\omega \omega' l}$ are Bogoliubov coefficients that transform between the black hole and 
 cosmological bases of states via the Klein-Gordon inner product. The cut-off frequency $\omega_0$, 
which is determined by the classical scattering of a scalar field in SdS, depends on both the size of the black hole
 and which horizon is the absorber.
 The length scale $R$ is the light-travel time between the emitting and absorbing horizons and arises
 from the use of properly normalized wave packets\footnote{ This amounts to ``putting the black hole in a box" and using
 discrete modes with frequencies $\omega_n = n/R$.}.

 We consider separately the limits of small and large black holes.
 In the small black hole limit, $r_b\ll r_c$, one finds that the cut-offs for black hole and cosmological particles to be given respectively by
 \begin{align}
 \label{cutoffsmall}
\omega_0  & \simeq T_b, \quad\quad\,\,\,\,\textrm{for}\  \N^b  \\
  \omega_0  & \simeq 2\pi l T_b^2, \quad \textrm{for}\  \N^c  
 \end{align}
 In the large black hole limit, with $r_b\simeq r_c\simeq l/\sqrt{3}$ and $T_b\simeq T_c\ll 1/l$, the emission and absorption processes for the two horizons become increasingly symmetrical and one finds
%
 \be\label{cutofflarge}
 \omega_0 \simeq (2\pi T) ^{3/2} l^{1/2} 
 \ee
where $T$ is the approximately common temperature of the two horizons.
%
For large black holes, it is found  the number of particles produced per unit time at frequency $\omega$ is found in \cite{qiu} to be given by
\begin{equation}\label{Nlarge}
 \N^h
 \simeq{ \omega  \over 3 \pi^3 T } 
\exp\left( - {4\omega^{1/2} l^{1/4} \over  (2\pi T)^{1/4} } \right) d\omega
\end{equation} 
where the light travel time was found to be related to the common horizon temperature by
$R\simeq r_c -r_b \simeq 2\pi l^2\, T$  and the result applies to production from both the black hole and cosmological horizons. 
The large, cold black hole
 is a quasi-equilibrium state, with the two temperatures differing only  at second order in the small quantity $\mu$ in (\ref{sdstemps}).
Further, since the black hole and cosmological horizon areas are approximately equal, each horizon absorbs and emits equally to leading order.

However, Eq.~(\ref{Nlarge}) is not the full story.   There is an additional geometric consideration
affecting what modes may contribute to measurements made by a near-horizon observer.
The wavelength must be 
less than the propagation distance, or equivalently that $\omega>\omega_1$ with
\be\label{omegafit} 
\omega_1 \simeq {1\over 2\pi l^2 T}  
\ee
%
 This is a more stringent lower cutoff on $\omega$ than $\omega_0$ in (\ref{cutofflarge}), implying that contributing frequencies are in the 
high frequency  tail of the distribution of (\ref{Nlarge}).  The total number of particles emitted by each horizon per unit time,
 given by the integral of (\ref{Nlarge}) over frequencies larger that $\omega_1$, 
 is therefore quite small and given by
\begin{equation}\label{ntotbig} 
n_h \simeq {2\over 3\pi l}{1\over (2\pi lT)^{9/4} }  e^{- 4 /(2\pi l T)^{3/4}   } 
 \end{equation} 
 which again applies to emission from both the black hole and cosmological horizon in the large black hole limit.
 
The energy in scalar particles crossing the cosmological horizon is obtained by multiplying
$\N^b $ by a factor of $\omega$ an integrating over frequency, giving
\be\label{energylarge}
  E^\phi_b \simeq { 1 \over 48\pi^2  l} {1 \over (2\pi l T) ^{11/4} }  e^{- 4 /(2\pi l T)^{3/4}   } 
\ee
where we have 
multiplied by a factor of $R$ to give the energy rather than energy per unit time.
The corresponding entropy follows from the classical first law $dE=TdS$
applied to the produced particles\footnote{The first law implies that that the temperature dependence of the energy and entropy of the radiated particles should be related according to ${dE\over dT}= T{dS\over dT}$.}, which gives
\be\label{entropylarge}
S^\phi_b  
\simeq\ { 1 \over 48\pi^2  l T} {1 \over (2\pi l T) ^{11/4} }  e^{- 4 /(2\pi l T)^{3/4}   } 
\ee
Note that both the energy and entropy are exponentially suppressed in this low temperature regime.

To summarize, as the common temperature of the black hole and cosmological horizons goes to zero in the large black hole limit,  each of the  $\N^h$ for $h=b,c$ tends to zero, with the
 suppression coming from the fact that only frequencies
 in the tail of the Boltzmann distribution can contribute. 
(This is analogous to quantum freezeout in statistical systems at temperatures
much lower than the energy gap.) As a result,
  $n_h , E^\phi_b$, and $S^\phi_b$ and their derivatives with respect to $T$ go to zero.

In the small black hole limit, $r_b\ll r_c$, the black hole horizon temperature is much greater than the cosmological horizon temperature, with the expressions for the temperatures (\ref{temperatures}) given approximately by  
their values in Schwarzschild and deSitter respectively 
\begin{equation}
T_b \simeq {1\over 8\pi M},\qquad T_c \simeq {1\over 2\pi l }
\end{equation}
If we define the small parameter $\epsilon = 1/(l T_b)$, the particle production calculation shows that
through terms of order $\epsilon$ the black hole emission rate reduces to the Schwarzschild result, which can be written as
\begin{equation}\label{Nbrhosmall}
\N ^b = 
{  \omega \over  2\pi^2 T_b   }e^{-\omega/ T_b }\, d\omega 
\end{equation} 
where we in this limit we have set $R=l$. The total rate of particles crossing the cosmological
horizon, given by the integral over frequency of (\ref{Nbrhosmall}) with the low-frequency cutoff $\omega_0$ in (\ref{cutoffsmall}), is found to be
\begin{equation}\label{ntotbsmall}
n_b
\simeq {  T_b  \over 2\pi^2 } 
\end{equation}
The total rate of energy in particles crossing the cosmological horizon is obtained by multiplying
$\N^b$ by a factor of $\omega$ and then integrating, which yields
\be\label{energysmallb}
E^\phi_b = { l\,T_b^2 \over 4\pi^4 }
\ee
This differs from the usual  temperature raised to the fourth power dependence for the energy density of black body radiation, because it includes a factor of the black hole horizon area which scales as $1/T_b^2$ for small SdS black holes.
The entropy in the $\phi$-particles can again be determined via the first law, giving
%
\be\label{entropysmallb}
S^\phi_b  = {lT_b\over 3\pi^4}
\ee
The energy and entropy carried by particles radiated from the black hole diverge at high temperatures,
reflecting the thermal instability of small Schwarzschild black holes 
which persists with $\Lambda>0$. 
However, a small black hole carries only a small 
amount of energy and therefore this emission cuts off after only a short amount of time due to back-reaction.   
In order to make contact with the Schottky anomaly, this back-reaction must be taken into account. If we assume that the black hole mass is reduced by the energy produced in $\phi$-particles, we have that 
 \be\label{mloss}
 {dM\over dU_c } = -{1 \over l}E^\phi_b
 \ee
 %
where $U_c$ is the null Kruskal coordinate on the future cosmological horizon\footnote{For an asymptotically black hole, this would be replaced by a good coordinate at future null infinity.}. 
 Next, we observe that since the black hole temperature is approximately
 $T_b \simeq 1 /8\pi M$, that we can rewrite this in terms of the change in black hole temperature with respect to Kruskal time.  Making use additionally of (\ref{energysmallb}), we have
 %
 \be\label{dudt}
{dT_b\over dU_c} = {2\over \pi^3}T_b^4
 \ee
 which is positive, since the black hole is heating up as it loses mass.
 The amount of energy $\Delta E^\phi_b$ emitted in $\phi$-particles in an interval $\Delta U_c$ of Kruskal time on the cosmological horizon is then related to the change $\Delta T_b$ in the black hole horizon temperature by
 \begin{align}\label{eradu1}
\Delta E^\phi_b &\equiv  {E^\phi_b \over l} \Delta U_c \\ &= {1 \over 8\pi T_b^2 } \Delta T_b\label{eradu}.
 \end{align}
 Integrating this expression returns the input that the total energy emitted is equal to the initial (small) mass of the black hole.
Similarly, the change in entropy $\Delta S^\phi_b$ emitted in $\phi$-particles in an interval $\Delta U_c$ of Kruskal time is related to 
$\Delta T_b$ by
 \begin{align}\label{sradu1}
 \Delta S^\phi_b &\equiv {S^\phi_b \over l} \Delta U_c \\ & \simeq  {1 \over 6\pi T_b^3 } \Delta T_b\label{sradu}
 \end{align}
The high temperature limit is complicated because of the many scales involved.  In addition to the distance $l$ between the 
horizons and the black hole temperature $T_b$, the Kruskal time interval $\Delta U_c$ over which the black hole mass decreases by a significant fraction of its initial value ({\it e.g.} $1/2$) is a measure of the out-of-equilibrium nature of the system.
The expressions for the
emitted energy and entropy  (\ref{eradu}) and (\ref{sradu}) are the same as
for asymptotically flat black holes. The difference between the Schwarzschild and SdS cases is contained in the definitions (\ref{eradu1}) and (\ref{sradu1}) which specify that the radiated energy and entropy are crossing
 the cosmological horizon within an affine time interval $\Delta U_c$.  Recall that particles are interpreted as emission from the black hole,
providing the flux at the cosmological horizon.
The analogous statements with $\Lambda=0$ are with respect 
 to the outgoing null coordinate at future null infinity, or at some arbitrarily chosen, very large sphere.

%

We now turn to particle
production associated with the cosmological horizon and absorbed by the black hole.  This turns out to be negligible in the small black hole limit, when compared to
the effects of black hole particle production.
Making use of the low frequency cutoff $\omega_0$ in (\ref{cutoffsmall}) one finds that the rate of particle production from the cosmological horizon is given by\footnote{When doing the integral
to get $n^c$ from $N^c_\omega$ one must use the exact expression for $y(\omega ) $, given by 
$ y = {2\pi\over \epsilon } \omega^{\frac{\epsilon}{1+\epsilon}}(\omega_0)^{\frac{1}{1+\epsilon}}$. Indeed, it is 
most straightforward to use the general form for $N^c$, do the integral, and then take the limit of 
$\epsilon \ll 1$.} 
 %
%
\begin{align}\label{Ncrhosmall}
\N^c &= { \omega A_b \over 4\pi^3  l }  e^{-y}\, d\omega \\
& = { \omega \over 4\pi^4  l T_b^2 }  e^{-y} \, d\omega 
\end{align}
where  $y(\omega )={1 \over \epsilon } \left(  \epsilon l \omega \right)^\epsilon$ and $\epsilon \equiv T_c / T_b  \ll 1$.
The total number of particles emitted per unit time is found to be
\begin{equation}\label{ncsmall} 
n_c  \simeq {1 \over 2\pi^4 (\pi-1) l } e^{-4\pi^2 l T_b }  
\end{equation}
We see therefore that the quantities $\N^b$, $\N^c$, and $n_c$ all tend to zero in the limit that the black hole temperature $T_b$ goes to infinity. 
One can further check that the
derivatives of these quantities with respect to $T_b$ also tend to zero in this limit. 
We have found in this limit that the suppression of $\N^b$ and $\N^c$ comes from the 
density of states and the lifetime factors, which are made small by the small area of the black hole horizon, whereas in the low
$T_b$ limit suppression comes from the Boltzmann factor.
Since production of cosmological
particles is small compared to black hole emission, the entropy and energy of produced particles is
dominated by those crossing the cosmological horizon. This is in accord with the results of the first law analysis 
 for classical fluctuations in (\ref{stotsmall}), which showed for small black holes that the change in the total entropy is primarily
 due to the change in the area of the cosmological horizon, but is driven by the Schwarzschild black hole physics.

The particle production described above reflects an intricate interaction between the black hole and cosmological horizons. A  mode that originates near $\hb$, and has positive frequency 
with respect to the local geodesic time coordinate $U_b$, will partly propagate 
through $\hc$ and partly scatter back into the black hole. Not all of the wave that is transmitted to $\hc$
 contributes to particle excitations. Near $\hc$
it divides into positive and negative frequency portions defined with respect to the local geodesic time coordinate
$U_c$ on $\hc$, which differs from $U_b$. 
The same considerations apply to modes that start near the cosmological horizon and are either absorbed by
$\hb$ or pass out through $\hc$. 

It is interesting to observe aspects of the radiation processes that are reflective of the Schottky anomaly. In analogy with the  spin models, we may think of 
the production of particles from either the black hole horizon $\hb$ or the cosmological horizon $\hc$ corresponds to
excitations to  higher energy states.  The degree of this excitation is limited by the bounds
on the mass and entropy of the SdS system. The particle production calculation is, of course, more complicated than the simple 
paramagnet discussed earlier, in which  some spins will flip with some probabilities
following the addition of a bit of energy to the system. 
For the modes of a quantum field in SdS, the flip is the result of both wave propagation and a transformation between
two infinite dimensional bases of states. One could try to make the model of the paramagnet more detailed by adding interactions that mediate the spin flips, making the statistical mechanics system more similar to the black hole in its details.

As noted above, the particle production results
for small SdS black holes reduce in form to those for $\Lambda =0$: specifically, the quantities $ \Delta E^\phi_b$ and
$  \Delta S^\phi_b $ vanish in the limit of large $T_b$.  Not surprisingly, the Schwarzschild and SdS results differ
most significantly at low temperatures. With $\Lambda =0$, Eqs.~(\ref{eradu}) and~(\ref{sradu}) for the emitted energy
and entropy apply for all values of the black hole temperature $T_b$. They describe monotonic curves which diverge as $T_b$ goes to
zero, as the geometrical area and mass do in this case as well. 
With $\Lambda >0$ the large cold black holes have exponentially
suppressed emission of energy and energy, given in Eqs.~(\ref{energylarge}) and (\ref{entropylarge}). 
Hence the energy and entropy production is maximal at some intermediate temperature, reflecting Schottky anomalous behavior in this temperature range.

\section{Conclusions}\label{conclusions}

The Schottky anomaly  identified in Section (\ref{twolevelsec}) is a new feature of Schwarzschild-deSitter black holes that is not present for Schwarzschild or  Schwarzschild-Anti-deSitter black holes. Its existence is tied to the presence of the cosmological event horizon.  As noted in \cite{Johnson:2019vqf} the energy scale determined by the location of the Schottky peak is that of the underlying microscopic degrees of freedom of the black hole.  We have discussed how the non-equilibrium character of SdS black holes can be captured by considering constrained states of a heat bath representing the deSitter background.  A simple paramagnetic model of microstates can realize some of the features of SdS black hole thermodynamics.  To do better, one might need to consider more refined or interacting models \cite{Banks:2006rx,Cotler:2016fpe} such as those used to model the fast scrambling properties of black holes \cite{Sekino:2008he}.  Finally, we have reviewed the results of recent calculations \cite{qiu} of particle production in SdS spacetimes in the context of the Schottky anomaly and the spin system model.


It will be interesting to explore the Schottky anomaly in larger families of deSitter black holes.  It is possible that including charge and angular momentum may introduce further new features.  For example, with nonzero charge there will be a maximum temperature for the black hole horizon, which for sufficiently large charge may remove the Schottky peak entirely.
We can also look for peaked behavior in the thermodynamic observables of 
other multi-horizon black hole spacetimes and perhaps view such behavior as a window on underlying quantum degrees of freedom.  For example, one could study the region between inner and outer horizons of asymptotically flat Reissner-Nordstrom or Kerr black holes, potentially making contact with other ``observed" phenomena such as the product rule for inner and outer horizon entropies \cite{Larsen:1997ge,Cvetic:2010mn}.  It is worth noting that the product rule was originally found in the context of a model of microscopic black hole degrees of freedom \cite{Larsen:1997ge}, and its exploration in a wider context was viewed as potentially providing a `looking glass' into the microscopics of more general black hole solutions \cite{Cvetic:2010mn}.  Further extensions  include higher dimensions, different horizon geometries, and to higher curvature theories of gravity.

{\bf Acknowledgements:}  We thank Chang Liu, David Lowe and Jon Machta for stimulating conversations.  PD is supported by NSF grant PHY-1719642.  YQ is supported by NSF grant PHY-1820675.  This work was performed in part at Aspen Center for Physics, which is supported by National Science Foundation grant PHY-1607611.  This work was completed at KITP and supported in part by the National Science Foundation under Grant No. NSF PHY-1748958.

\appendix

%

\end{document}